\documentclass[pre,aps,twocolumn]{revtex4}
\usepackage{graphicx}

\newcommand{\beq}{\begin{equation}}
\newcommand{\eeq}{\end{equation}}
\newcommand{\ba}{\begin{array}}
\newcommand{\ea}{\end{array}}
\newcommand{\bean}{\begin{eqnarray*}}
\newcommand{\eean}{\end{eqnarray}}
\newcommand{\bea}{\begin{eqnarray}}
\newcommand{\eea}{\end{eqnarray}}
\newcommand{\bc}{\begin{center}}
\newcommand{\ec}{\end{center}}
\newcommand{\bt}{\begin{table}}
\newcommand{\et}{\end{table}}

\newcommand{\beqno}{\begin{displaymath}}
\newcommand{\eeqno}{\end{displaymath}}

\newcommand{\been}{\begin{enumerate}}
\newcommand{\een}{\end{enumerate}}

\begin{document}

\title{X,Y,Z-Waves: Extended Structures in Nonlinear Lattices}
\author{P. G.\ Kevrekidis$^{1}$, J. Gagnon$^{1}$, D. J.\ Frantzeskakis$^{2}$
and B. A.\ Malomed$^{3}$}
\affiliation{$^{1}$ Department of Mathematics and Statistics, University of
Massachusetts, Amherst MA 01003-4515\\
$^2$ Department of Physics, University of Athens, Panepistimiopolis,
Zografos, Athens 15784, Greece \\
$^3$ Department of Interdisciplinary Studies, Faculty of Engineering, Tel
Aviv University, Tel Aviv 69978, Israel}

\begin{abstract}
Motivated by recent experimental and theoretical results on
optical X-waves, we propose a new type of waveforms in 
%two- and three-dimensional 
2D and 3D discrete media -- multi-legged extended
nonlinear structures (ENS), built as arrays of lattice solitons
({\it tiles} or {\it stones}, in the 2D and 3D cases,
respectively). First, we study the stability of the tiles and
stones analytically, and then extend them numerically to complete
ENS forms 
%. Results are reported 
for both 2D and 3D lattices.
%including the first analytical stability calculations in the latter case. 
The predicted 
%extended 
patterns are relevant to a variety of physical settings, such as Bose-Einstein condensates in
deep optical lattices, lattices built of microresonators, 
photorefractive crystals with optically induced lattices 
(in the 2D case) 
and others.
\end{abstract}

\maketitle

\textit{Introduction and setup}. In the past few years, studies of dynamics
in discrete media, as well as in their continuum periodically modulated
counterparts, 
have enjoyed very rapid progress \cite{general_review}. 
This is fueled by a need to
tackle diverse physical contexts such as nonlinear optics of waveguide
arrays and photorefractive crystals \cite{review_opt}, the physics of
Bose-Einstein condensates (BECs) in optical lattices \cite{review_BEC}, 
%and
the denaturation of the DNA double-strand \cite{review_DNA}, and so on.
%among many others \cite{general_review}. 
These efforts have resulted in the creation of
a variety of novel nonlinear structures, such as discrete dipole \cite{dip},
quadrupole \cite{quad}, necklace \cite{neck} and other
multi-pulse/multi-pole localized patterns \cite{multi}, discrete vortices 
\cite{vortex}, 
%and 
ring solitons \cite{rings}, and others \cite{us_yuri}.
Such structures have a potential to be used as carriers and conduits for
data transmission and processing, in the context of all-optical 
and quantum-information schemes.

The present work is motivated by the above-mentioned achievements, 
%in studies of the lattice solitons, 
as well as by recent 
developments in studies of \textit{X-waves}, which are
quasi-linear extended structures generated in three-dimensions (3D) \cite{ditrapani},
two-dimensions (2D) \cite{boris}, and even quasi-discrete \cite{hiza} media. These
waves arise when the linear second-order differential operator in
the 
%corresponding 
relevant wave equation is genuinely, or effectively \cite{ditrapani2},  
sign-indefinite (roughly speaking, a D'Alembertian instead of the Laplacian). 
A similar shape is featured by cross configurations formed by two intersecting domain
walls in a system of two 
%nonlinearity 
coupled 2D Gross-Pitaevskii equations \cite{cross}, with a difference that these structure are
essentially nonlinear ones. 

In the present work, we are dealing with the sign-definite
discrete Laplacian (nevertheless, one can
easily induce the sign-indefiniteness in the discrete wave
equation, applying the well-known \textit{staggering
transformation} \cite{general_review} along one coordinate only).
Instead of emulating the mathematical structure 
of the continuous wave equations supporting X-waves in the lattice medium, 
our aim is to construct stable extended nonlinear structures (ENSs) in \emph{discrete} media.
These wave structures are partly delocalized, by their nature,
ranging from a few sites to infinitely extended ones. An essential
difference between the quasi-linear X-waves in continua and the lattice
entities presented here 
is that the latter are \emph{strongly nonlinear} structures, being built
as arrays of individual lattice solitons (``tiles"); in that
sense, they somewhat resemble weakly localized 
\textit{hypersolitons} \cite{hm} and 
\textit{supervortices} \cite{HSakaguchi}, constructed as finite ensembles
of individual solitons. 
%in continuum models combining cubic or saturable nonlinearity and a square lattice). 
In fact, we initialize our analysis (see below) from a purely nonlinear
(\textit{anti-continuum}, AC) limit, without linear intersite
coupling on the lattice. The eventual shape of the extended
lattice waves features a great variety, including X-waves,
Y-waves, Z-waves and other forms, provided that the underlying
core (``tile") is dynamically stable. Each species is stable in a
finite interval of values of the intersite coupling constant. 
%(for general reasons similar to those analyzed in Ref. \cite{dep}).
Some of those patterns may definitely be of interest (for
instance, as star-shaped waveguiding configurations) to
applications in both nonlinear-optical and matter-wave contexts,
where they can potentially be created. 
%be (and, to some extent, have been \cite{multi}) created.

The analysis is performed in terms of the 
%paradigmatic 
discrete nonlinear Schr{\"{o}}dinger (DNLS) equation in 2D or 3D,
%two or three dimensions,
%
\begin{equation}
i\dot{u}_{\mathbf{n}}=-\epsilon \left( \Delta _{2}u\right)
_{\mathbf{n}}-|u_{\mathbf{n}}|^{2}u_{\mathbf{n}},  
\label{DNLS}
\end{equation}
where $u_{\mathbf{n}}$ is a complex amplitude of the electromagnetic wave in
an optical waveguide array (in the 2D case)
\cite{review_opt,DNLS_opt}, or the BEC\ wave function at nodes of
a deep (2D or 3D) optical lattice \cite{DNLS_BEC}, $\mathbf{n}\
$being the vectorial lattice index, and $\Delta _{2} $ the
standard discrete Laplacian. Further, $\epsilon $ is the constant
of the intersite coupling, and the overdot stands for the
derivative with respect to the evolution variable; the latter is $z$ in optical
arrays, or $t$ in the BEC model (or in 
%. Another physical realization of both 3D and 2D DNLS equations with the evolution in $t$ is provided by 
a crystal built of microresonators \cite{photons}).

The presentation of our results on ENS solutions of Eq.
(\ref{DNLS}) proceeds as follows. First, 
in the AC limit ($\epsilon =0$), 
%in Eq. (\ref{DNLS})], 
we specify main types of the 
%above-mentioned 
``tiles'', of which the 
%extended structures 
ENSs are to be composed. Then, we numerically extend these compositions into full
ENS solutions for $\epsilon >0$. Finally, we summarize our
findings and discuss further problems.

\textit{Analytical results: tiles, stones, and their stability}. We
look for standing-wave solutions to Eq. (\ref{DNLS}), in the form
of $u_{\mathbf{n}}=\exp (i\Lambda t)\phi _{\mathbf{n}}$, with
$\phi _{\mathbf{n}}$ satisfying the 
%stationary 
equation,
\begin{equation}
f(\phi _{\mathbf{n}},\epsilon )=\Lambda \phi _{\mathbf{n}}-\epsilon \Delta
_{2}\phi _{\mathbf{n}}-|\phi _{\mathbf{n}}|^{2}\phi _{\mathbf{n}}=0.
\label{steady}
\end{equation}
%
%Linearization of 
Perturbation of Eq. (\ref{DNLS}) around the solutions of Eq. (\ref{steady}) 
%shows that a set of a small perturbation
%of the solution and its complex conjugate is a zero mode of the following
leads to the linearization operator,
\begin{eqnarray}
\mathcal{H}_{\mathbf{n}}^{(\epsilon )} = \left(
\begin{array}{cc}
\Lambda -2|\phi _{\mathbf{n}}|^{2} & -\phi _{\mathbf{n}}^{2} \\
-\bar{\phi}_{\mathbf{n}}^{2} & \Lambda -2|\phi _{\mathbf{n}}|^{2}\end{array}\right)
%\nonumber \\
-\epsilon \Delta _{2}\left(
\begin{array}{cc}
1 & 0 \\
0 & 1\end{array}\right) ,  
\label{oper}
\end{eqnarray}
the overbar standing for complex conjugate. By means of a rescaling, we
fix $\Lambda \equiv 1$, keeping $\epsilon $ as a free parameter.
%where $s_{n'} \phi_n=\phi_{n+n'}$ is the shift operator (written
%in 3d in Eq. (\ref{energy})).

In the AC limit of $\epsilon =0$, solutions to Eq. (\ref{steady})
are 
%obvious, 
$u_{\mathbf{n}}=re^{i\theta _{\mathbf{n}}}$, where
real amplitude $r $ is $0$ or $\sqrt{\Lambda }$, and $\theta _{\mathbf{n}}$ 
are arbitrary constants.\ To continue such a
solution to $\epsilon \neq 0$, the Lyapunov-Schmidt condition
needs to be satisfied \cite{dep}, \textit{viz}., the projection of
eigenvectors of the kernel of $\mathcal{H}_{\mathbf{n}}^{(0)}$
(i.e., zero modes of the operator) onto the system of stationary
equations should be null. This condition gives rise to a
solvability condition at\ each ``AC-filled" site, i.e., one with
$r\neq 0 $ in the AC limit,
\begin{equation}
-2ig_{\mathbf{n}}(\theta ,\epsilon )=\epsilon e^{-i\theta
_{\mathbf{n}}}\Delta _{2}\phi _{\mathbf{n}}-\epsilon e^{i\theta
_{\mathbf{n}}}\left( \Delta _{2}\bar{\phi}\right) _{\mathbf{n}}
\label{inter}
\end{equation}
(the factor of $-2i$ is introduced for convenience). 
%Which is quite important for our present purposes, 
Importantly, the eigenvalues $\gamma $ of
the Jacobian, $\mathcal{M}_{ij}=\partial g_{i}/\partial \theta
_{j}$, are intimately related to the leading-order approximation
for eigenvalues ($\lambda $) of the linearization around a
stationary solution, $\lambda =\pm \sqrt{2\gamma }$. We use this
relation, alongside a perturbative expansion in the solution,
$\phi _{\mathbf{n}}=\phi _{\mathbf{n}}^{(0)}+\epsilon \phi
_{\mathbf{n}}^{(1)}+\dots $, to derive leading-order bifurcation
conditions for any given configuration, and subsequently to find
the corresponding linear-stability eigenvalues.

2D configurations can be categorized by the number of their ``legs\textquotedblright\ 
%(i.e., quasi-1D lines of AC-filled sites it is composed of).
(i.e., the number of quasi-1D lines of excited sites it contains).
Simplest is the one-leg structure assembled of
two- or three-site tiles, 
%This case is particularly simple as it allows to borrow 
for which 1D stability results are applicable 
%from Ref. 
\cite{dep}; here, the resulting Jacobian is
\[
(\mathcal{M}_{1})_{i,j}=\left\{
\begin{array}{lcl}
\cos (\theta _{j+1}-\theta _{j})+\cos (\theta _{j-1}-\theta _{j}), 
\quad 
i=j, \\
-\cos (\theta _{j}-\theta _{i}), 
\quad  
i=j\pm 1, 
\\
0, 
 \quad  
|i-j|\geq 2,\end{array}\right.
\]
from which 
%important 
conclusions for the stability of ENSs may be drawn. In
particular, a fundamental stability condition extensively used in
our analysis is that the phase shift between fields at adjacent
sites must be $\pi $ \cite{todd,dep}. For the two-site tile
satisfying this condition, the stability eigenvalues are
\cite{dep} $\lambda =\sqrt{2\gamma }=\pm 2\sqrt{\epsilon }i$,
while for its three-site counterpart, $(+1,-1,+1)$, they are
$\lambda =\pm \sqrt{2\epsilon }i$ and $\lambda =\pm
\sqrt{6\epsilon }i$. The full one-leg configuration assembled of
these tiles is a chain, $(\dots ,-1,+1,-1,+1,-1,\dots )$,
hereafter termed configuration 1.

\begin{figure}[tbp]
\begin{center}
\hskip-0.2cm\includegraphics[height=6cm,width=6cm]{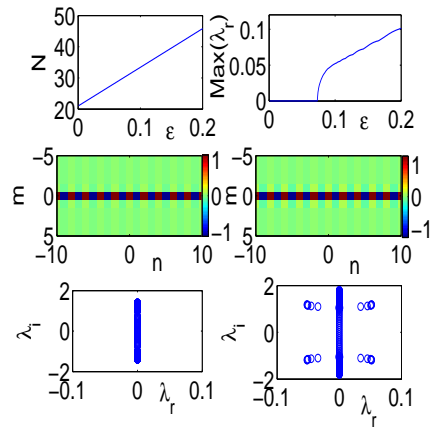} ~~
\end{center}
\par
\vskip-0.5cm \caption{(Color online) A family of single-leg
patterns in the 
%two-dimensional 
2D DNLS equation. The top left and
right panels show, respectively, the norm and the largest
instability growth rate vs. $\protect\epsilon $. The contour plots
in the middle panels show the stationary solutions  for
$\protect\epsilon =0.05$ (stable) and $\protect\epsilon =0.1$
(unstable).  The spectral planes $(\protect\lambda
_{r},\protect\lambda _{i}) $ of numerically computed eigenvalues
for these solutions are displayed in the bottom panels.}
\label{Fig1}
\end{figure}

\begin{figure}[tbp]
\begin{center}
\hskip-0.15cm
\begin{tabular}{cc}
\includegraphics[height=6cm,width=2.75cm]{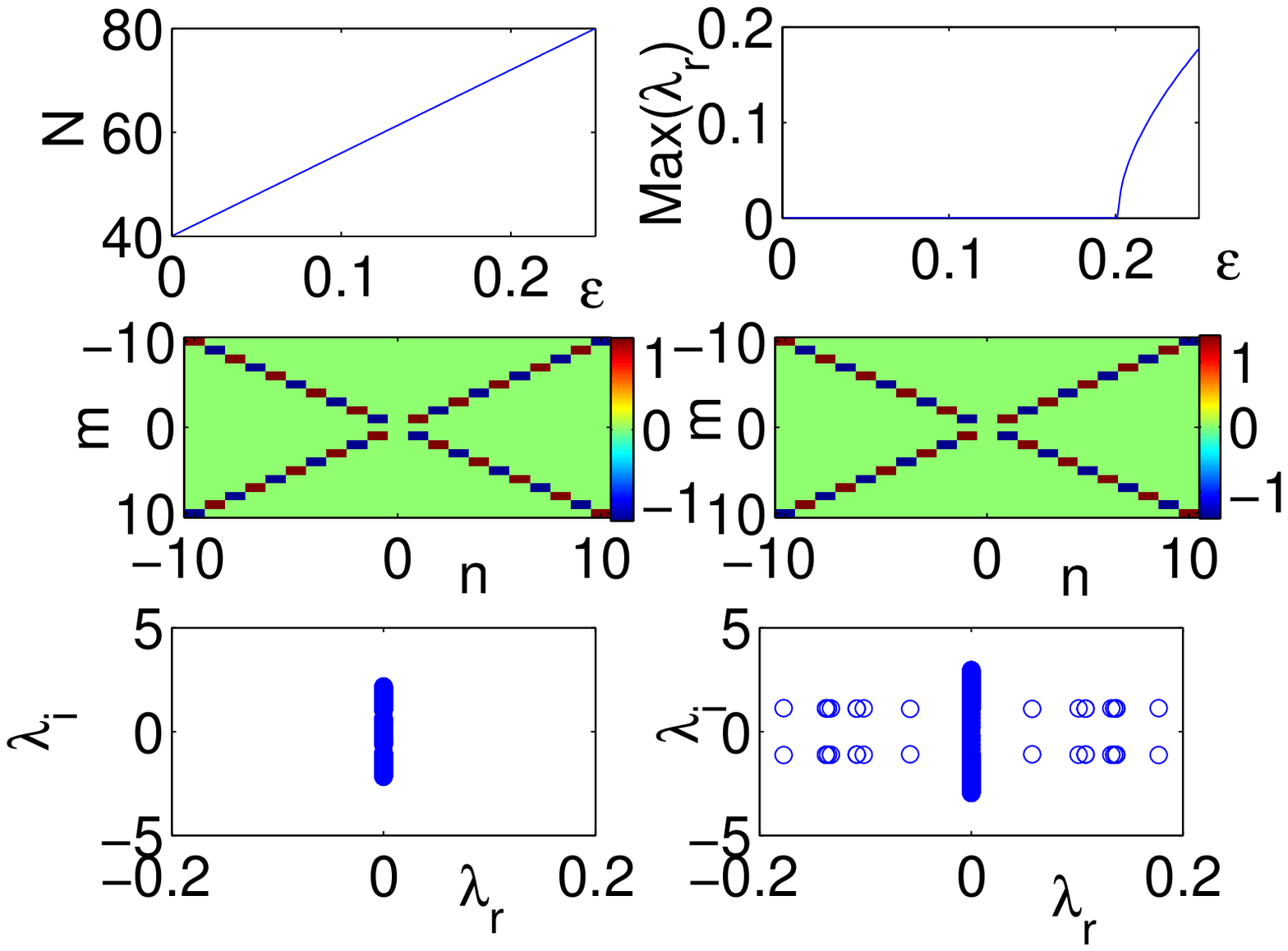}
\includegraphics[height=6cm,width=2.75cm]{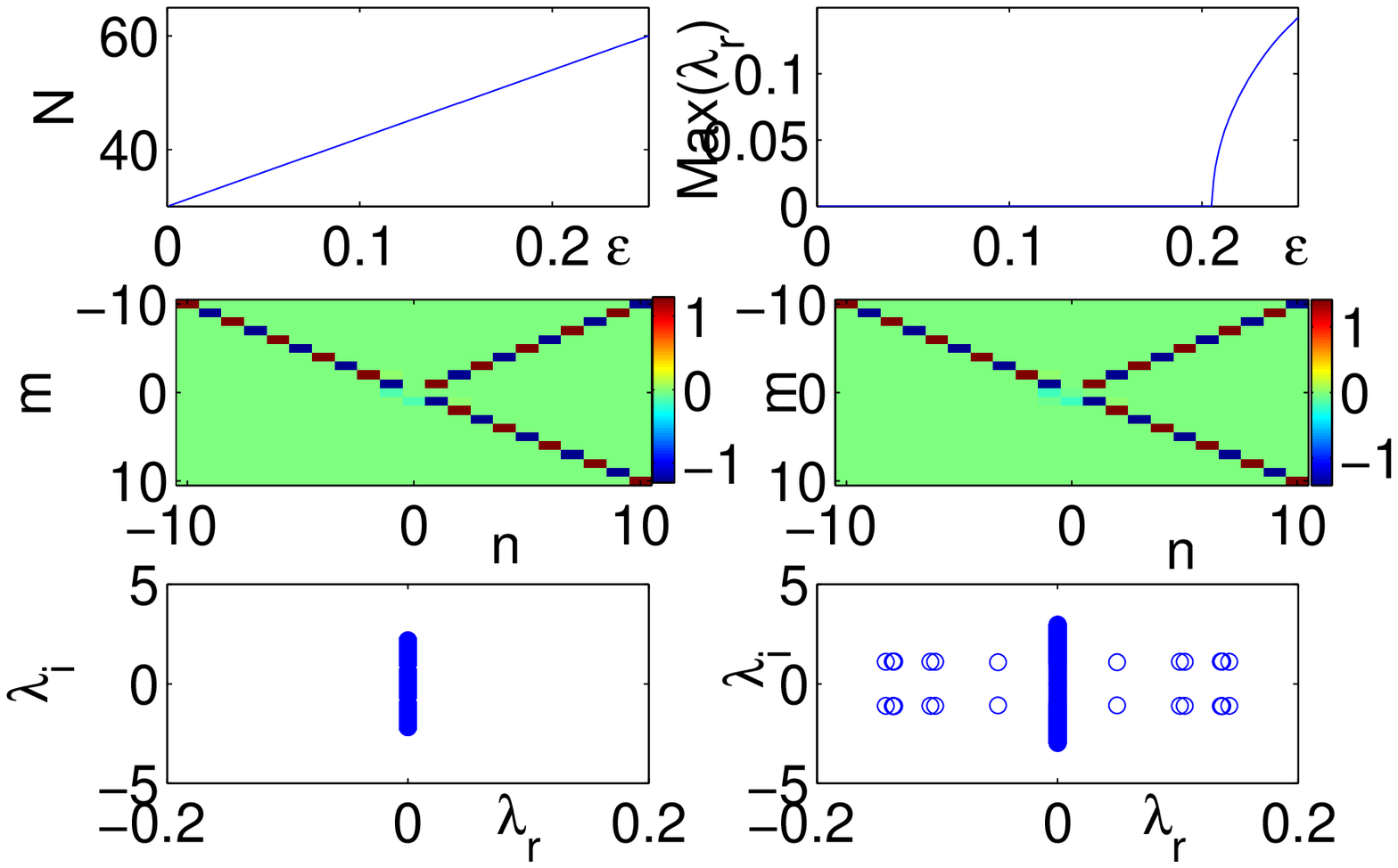}
\includegraphics[height=6cm,width=2.75cm]{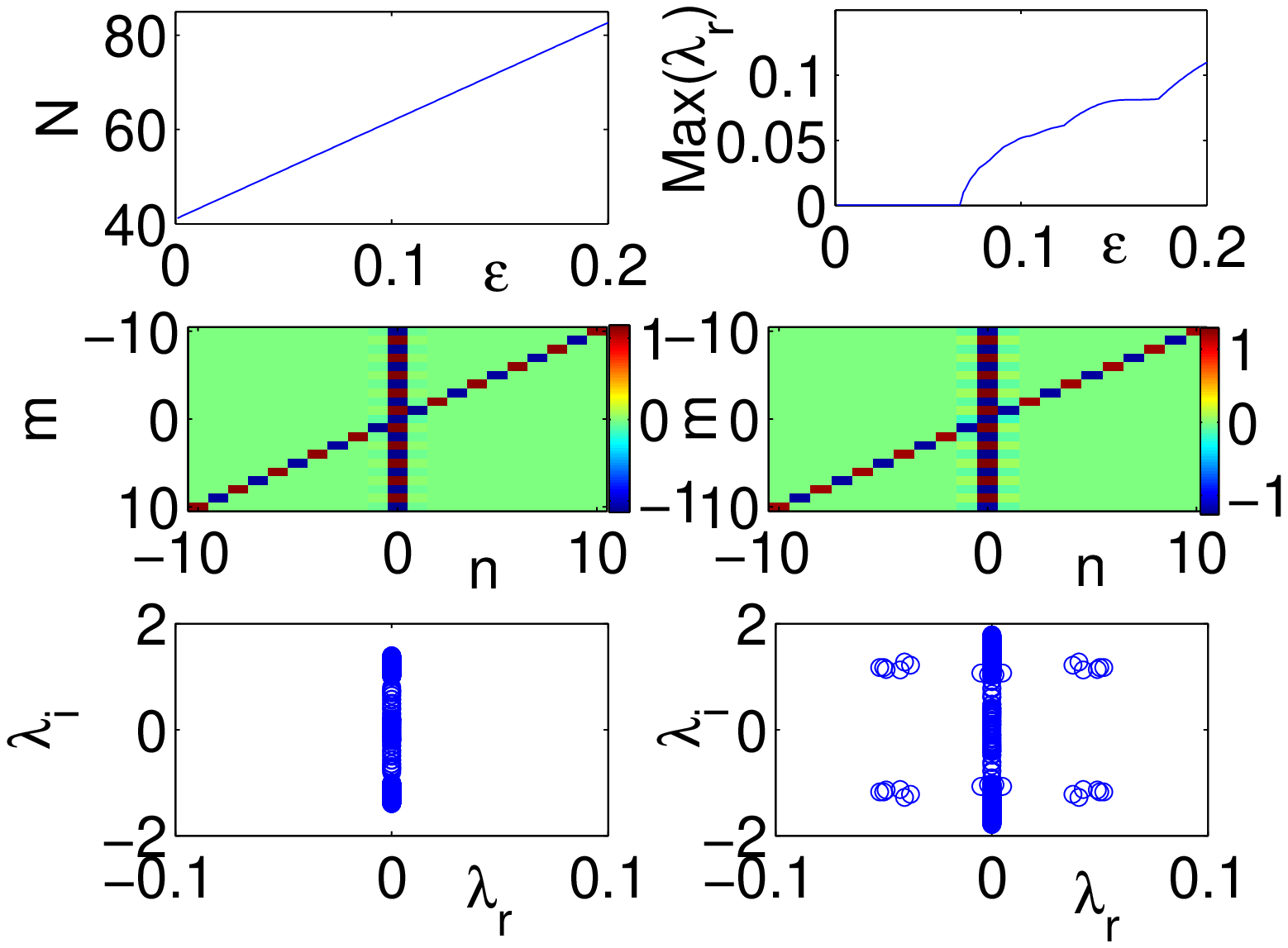} &  \\
\includegraphics[height=4cm,width=2.75cm]{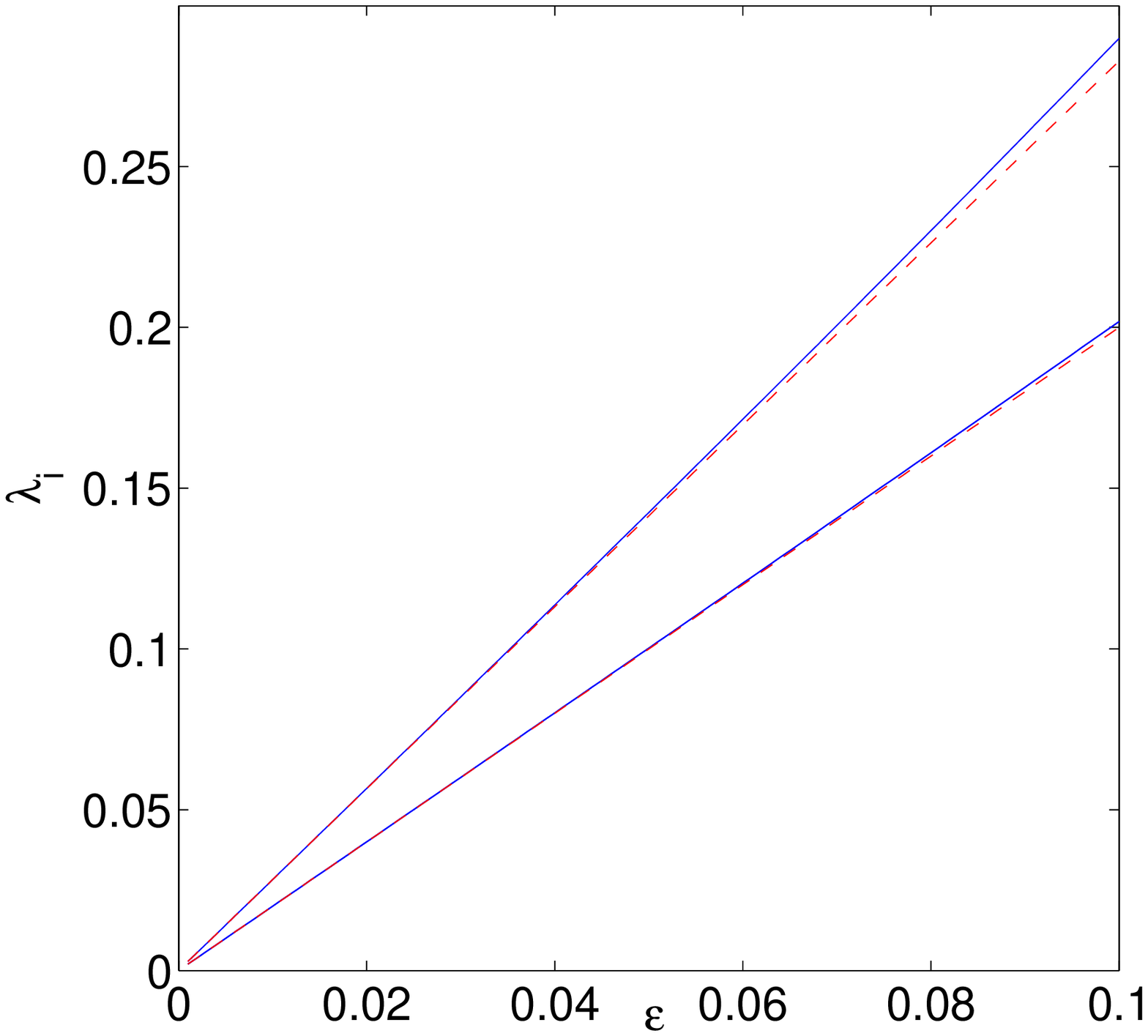}
\includegraphics[height=4cm,width=2.75cm]{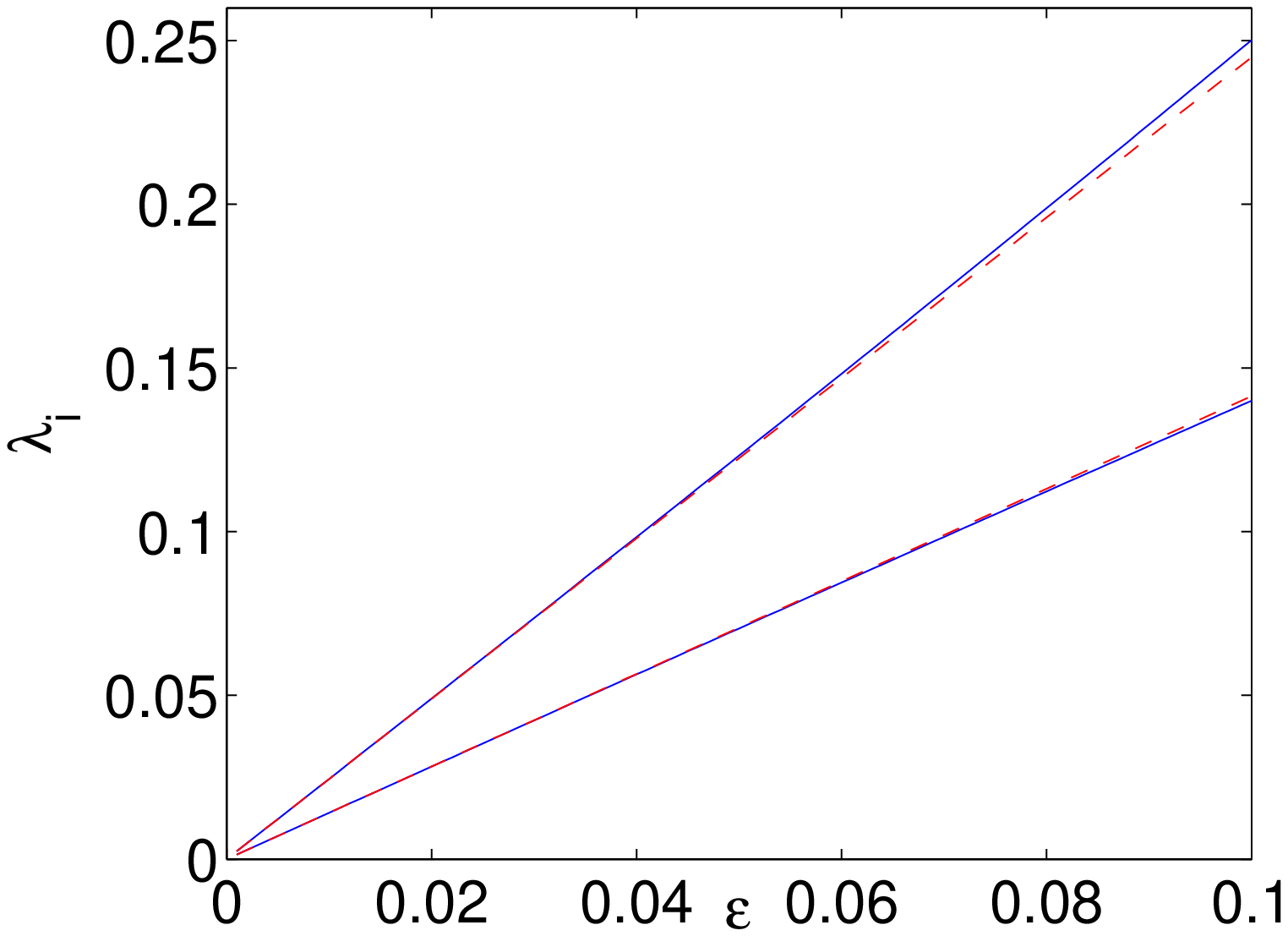}
\includegraphics[height=4cm,width=2.75cm]{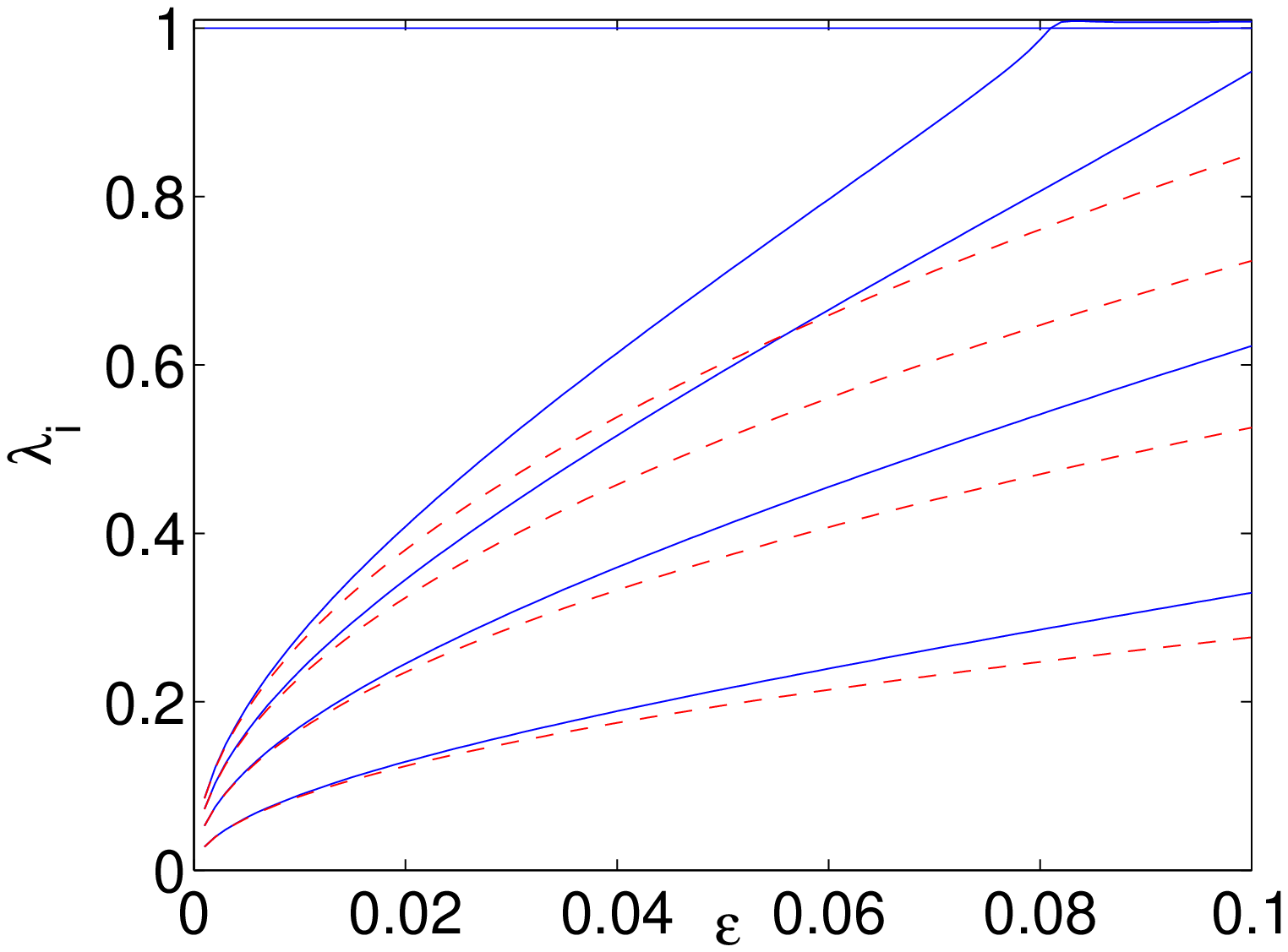} &
\end{tabular}\end{center}
\par
\vskip-0.7cm 
\caption{Same as in Fig. 1 but for configurations 2a, 2b and 2c. 
%Also included in 
Bottom panels present the stability
eigenvalues for individual tiles of which these patterns are
built, numerically found ones (solid 
%blue 
lines) versus the
analytical prediction (dashed 
%red 
lines). Examples of species 2a
and 2b are displayed in two left and two middle panels for
$\protect\epsilon =0.15$ and $0.25$, and an example of 2c in the
right panels pertain to $\protect\epsilon =0.05$ and
$\protect\epsilon =0.1$.} 
\label{Fig2}
\end{figure}

\begin{figure}[tbp]
\begin{center}
\hskip-0.15cm
\begin{tabular}{cc}
\includegraphics[height=6cm,width=4cm]{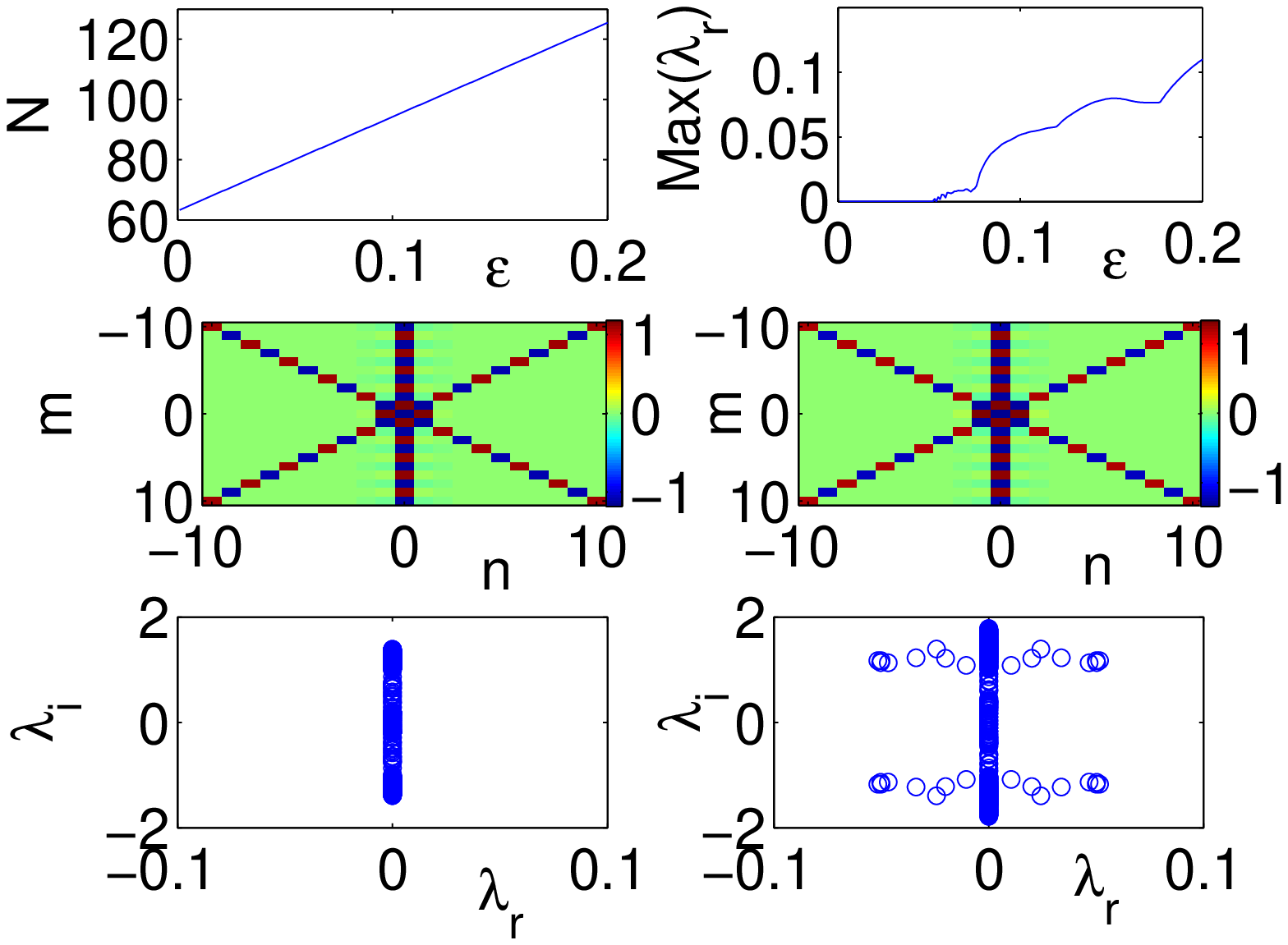}
\includegraphics[height=6cm,width=4cm]{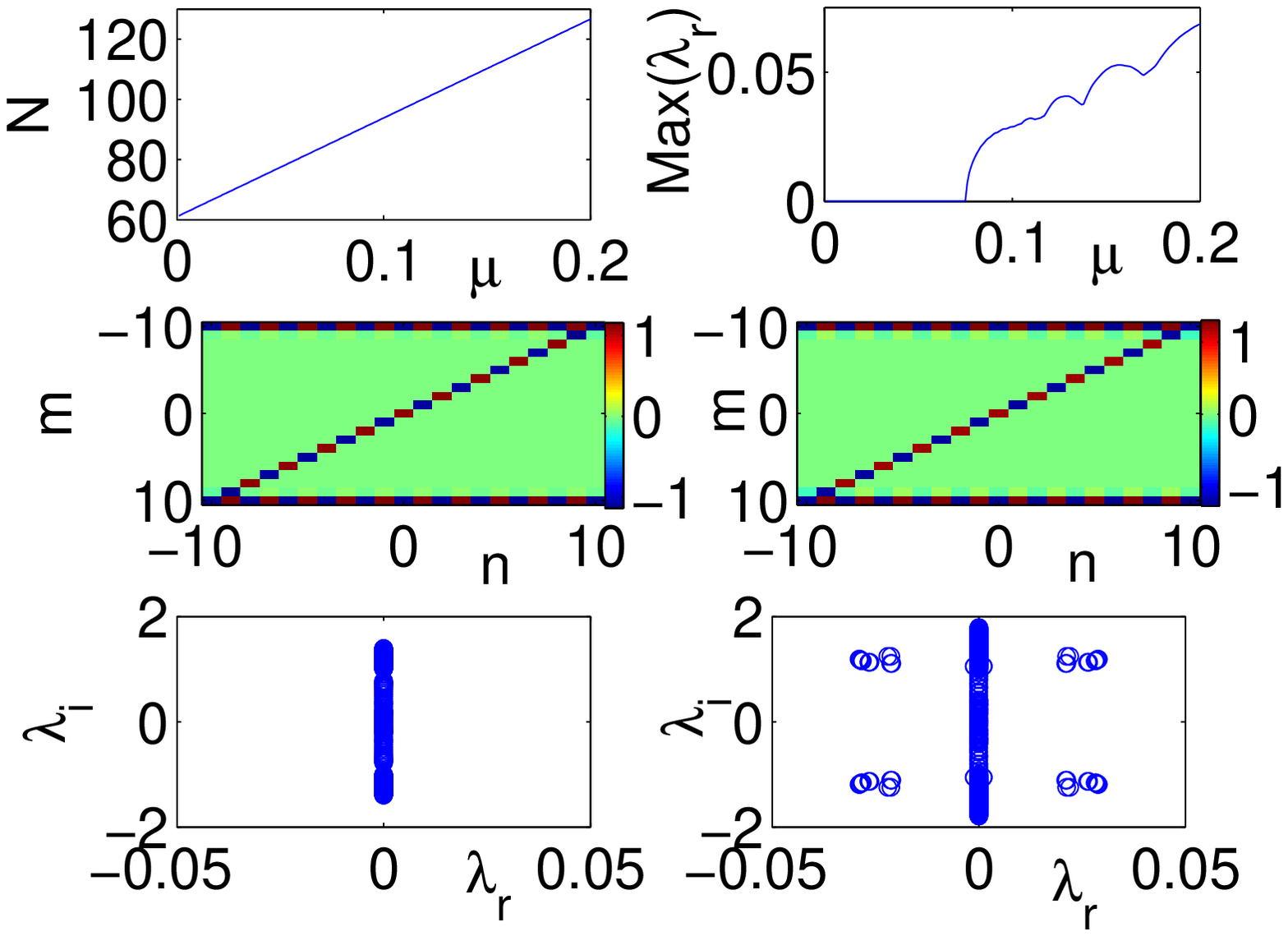} &  \\
\includegraphics[width=4cm]{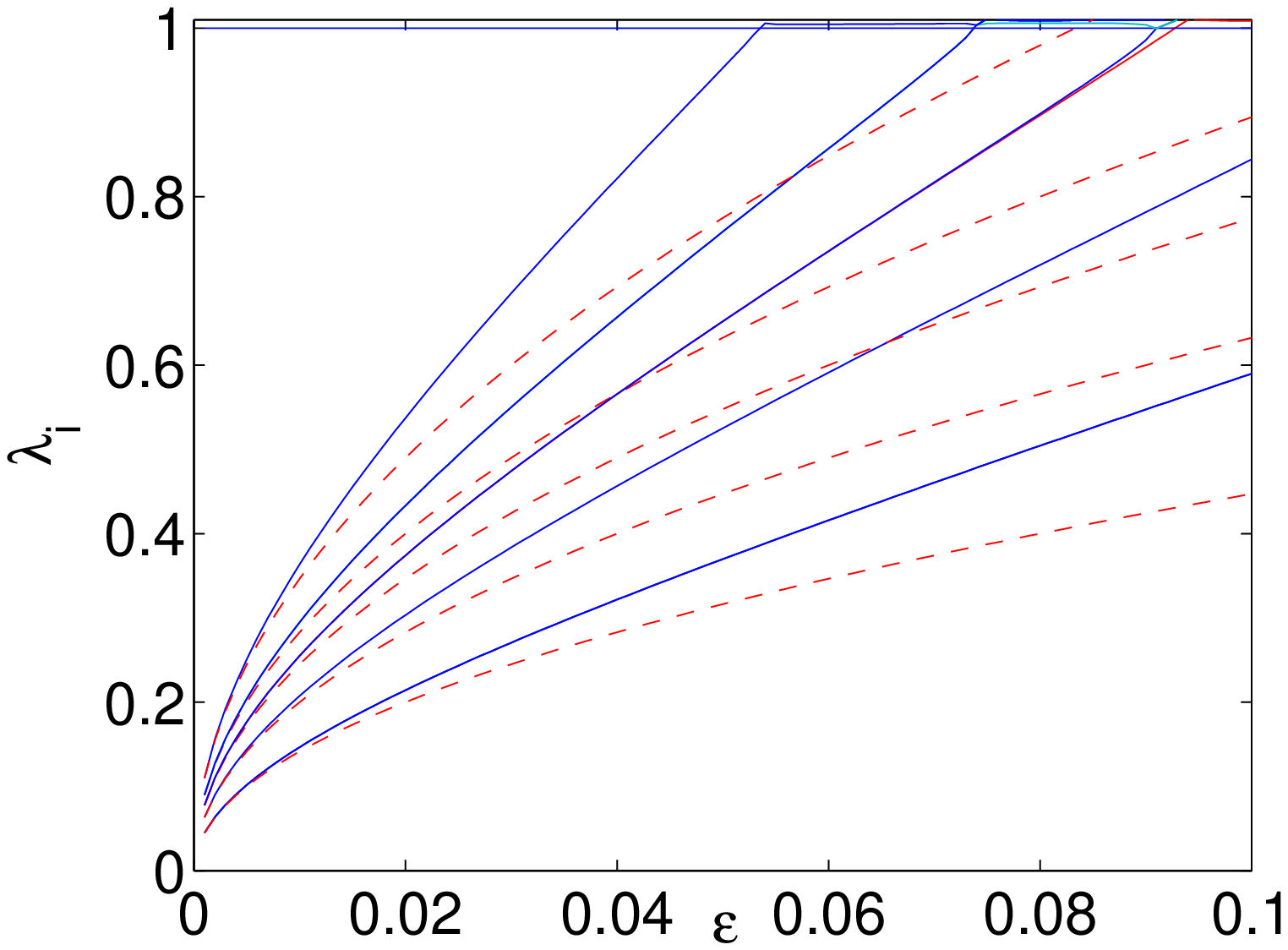}
\includegraphics[width=4cm]{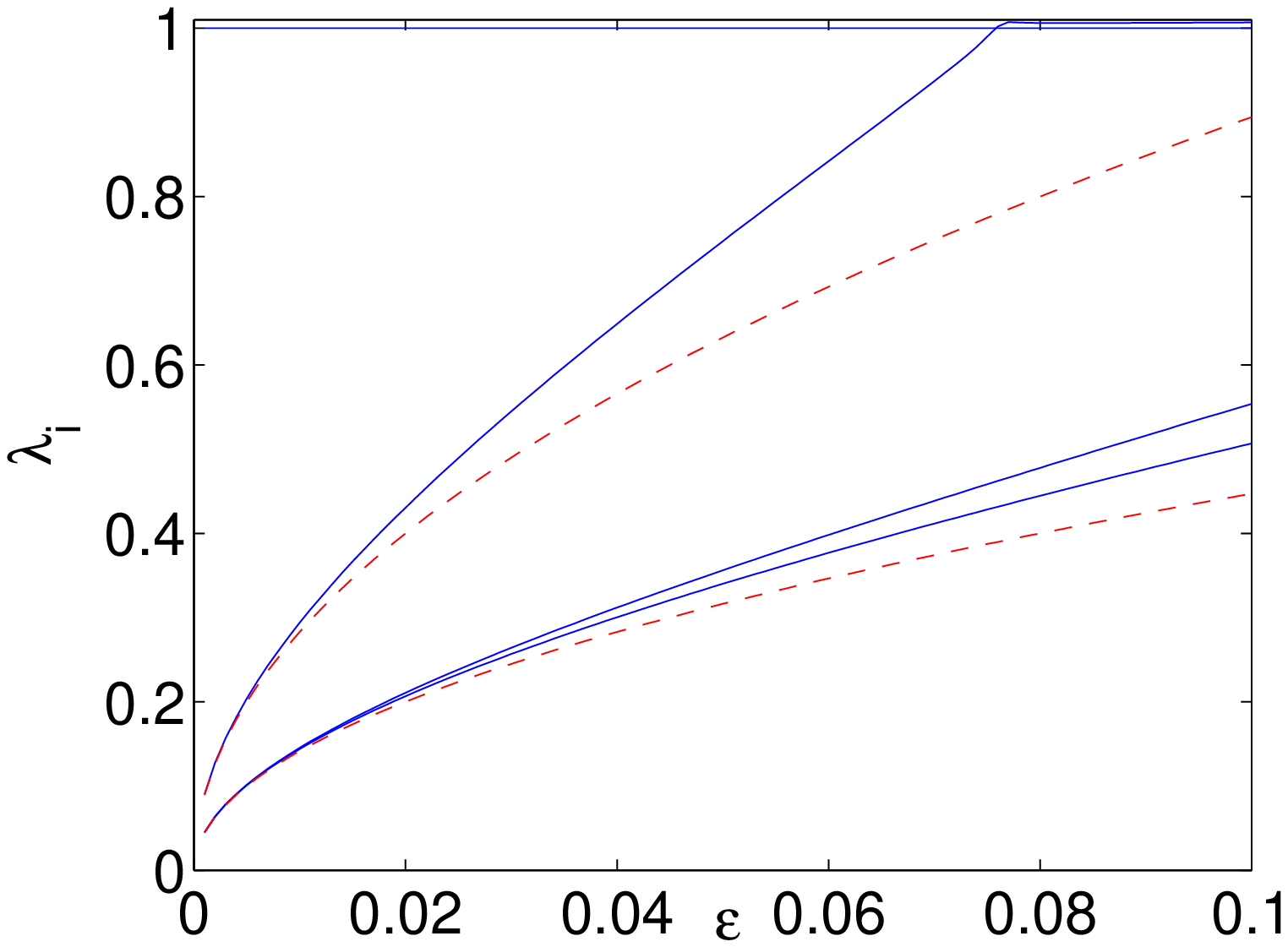} &
\end{tabular}\end{center}
\par
\vskip-0.7cm
\caption{Same as Fig. 2, but for configurations 3a and 3b. Examples of both
species are shown for $\protect\epsilon =0.05$ and $0.1$.}
\label{Fig3}
\end{figure}

\begin{figure}[tbp]
\begin{center}
\hskip-0.15cm
\begin{tabular}{cc}
\includegraphics[height=12cm,width=4cm]{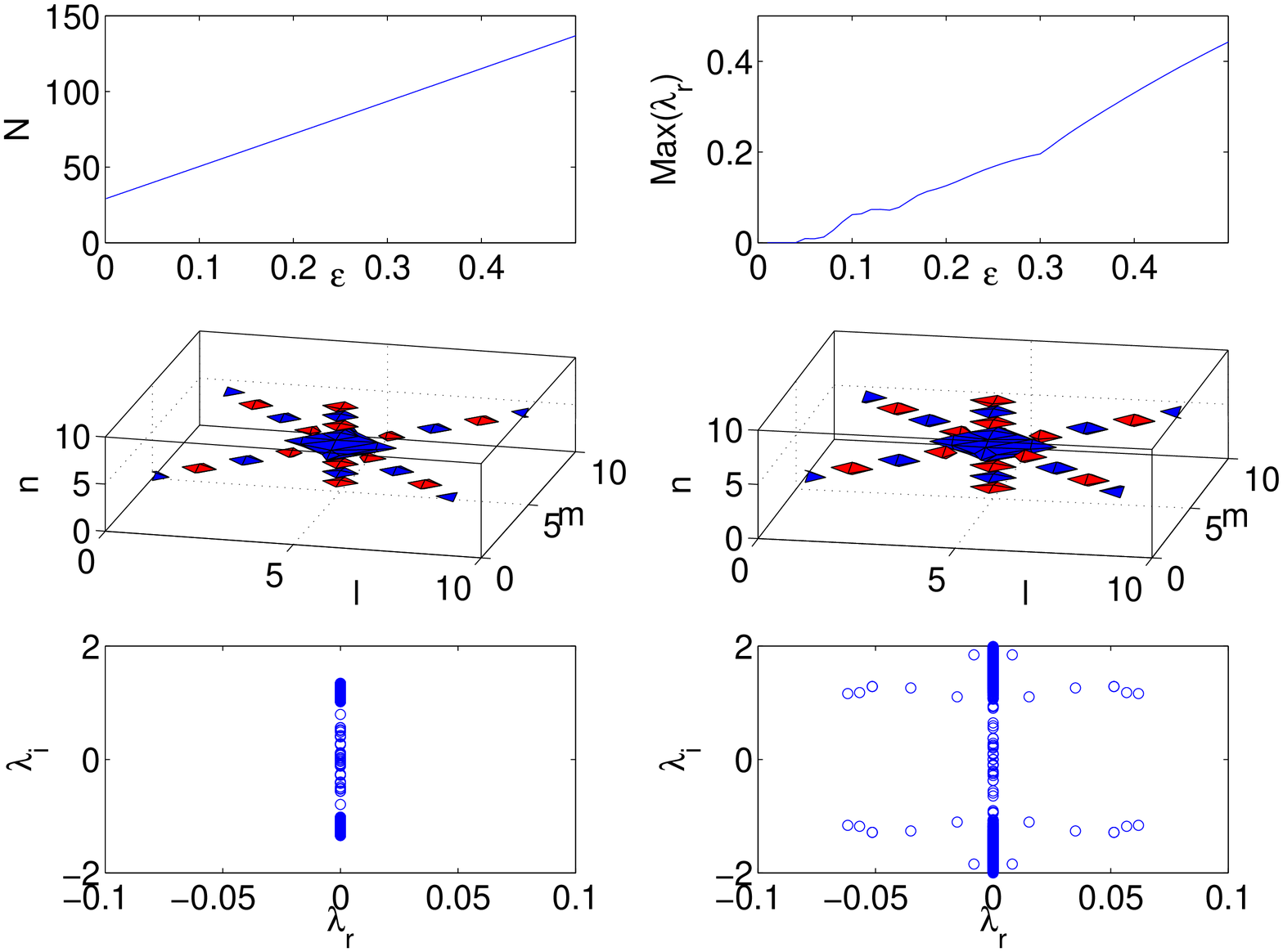}
\includegraphics[height=12cm,width=4cm]{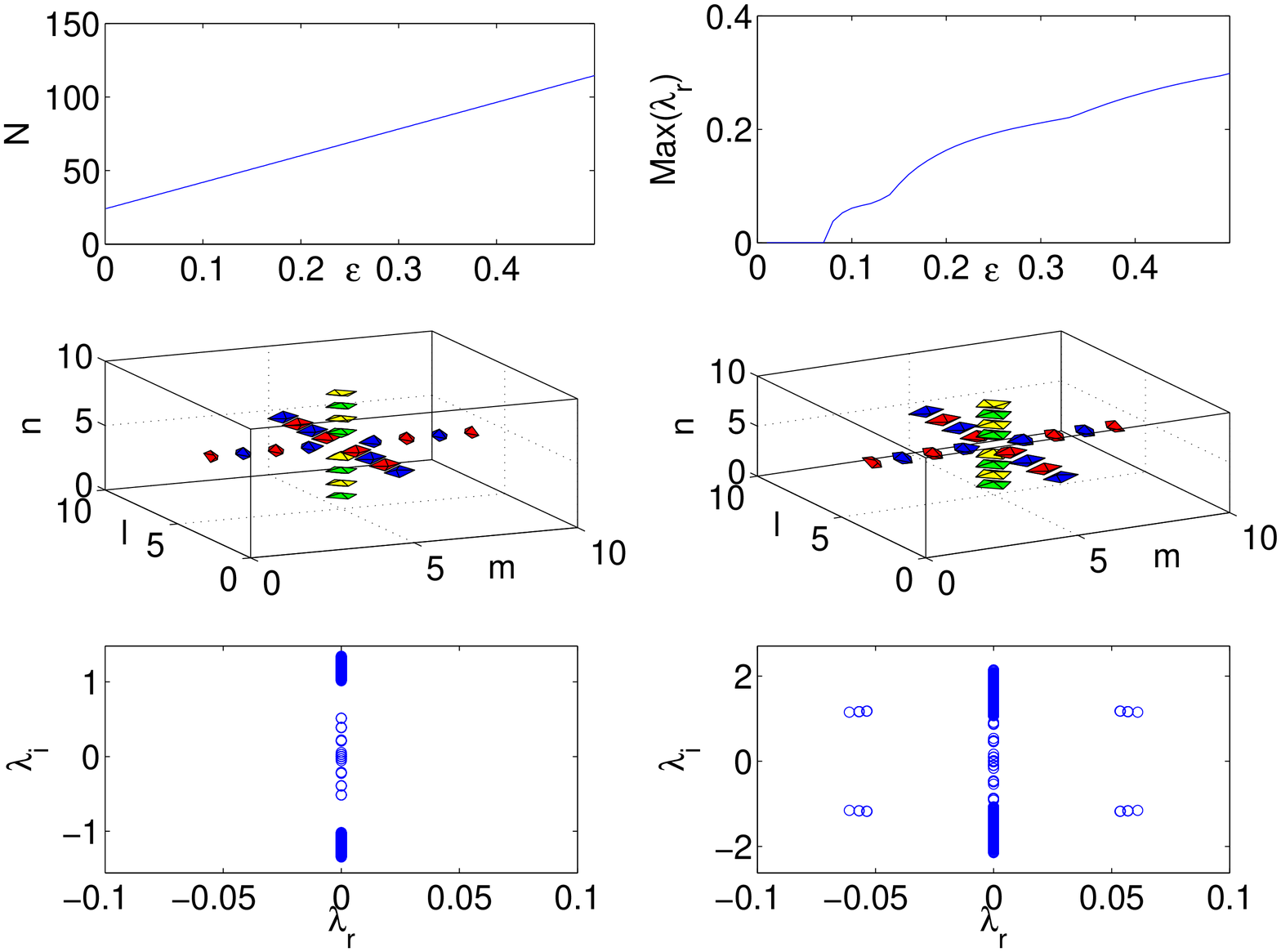} &  \\
\includegraphics[width=4cm]{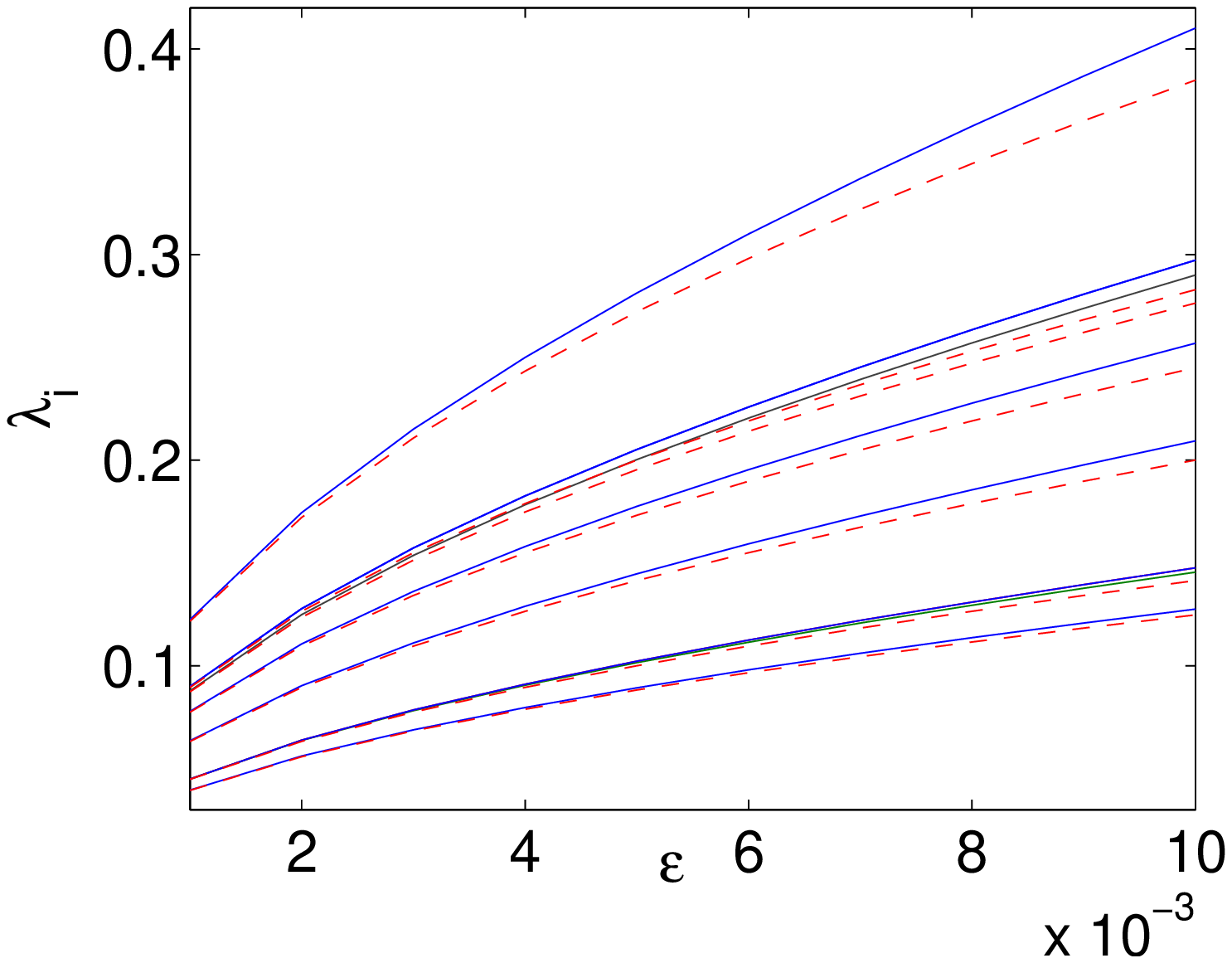}
\includegraphics[width=4cm]{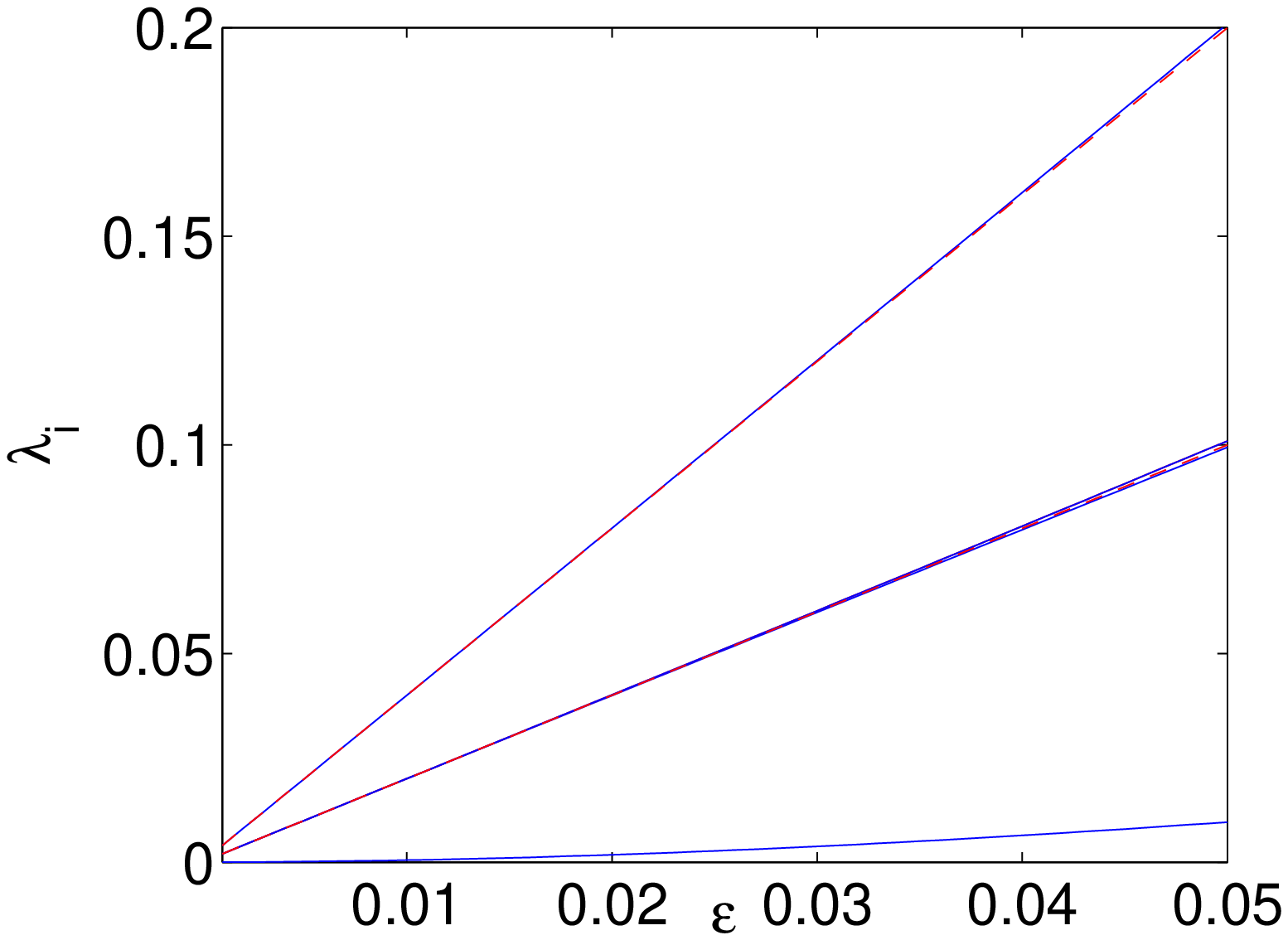} &
\end{tabular}\end{center}
\par
%\vskip-0.7cm
\caption{Same as in above figures, but for 3D configurations 4a
and 4b. Their examples are plotted, in two left and two right
panels, respectively, for $\protect\epsilon =0.03$ and
$\protect\epsilon =0.1$. In configuration 4b, the imaginary parts
are shown by different colors, green for phase $\protect\pi /2$
and yellow for $3\protect\pi /2$.} \label{Fig4}
\end{figure}

Proceeding to two-tile configurations allows us to examine more
complex (in particular, X- and Y-like) 2D structures.
X-configurations can be built of square- or cross-shaped tiles,
which, in the AC\ limit, are
\begin{equation} \left(
\begin{array}{ccc}
1 & 0 & -1 \\
0 & 0 & 0 \\
-1 & 0 & 1\end{array}\right) ~\mathrm{and~}\left(
\begin{array}{ccc}
1 & 0 & 1 \\
0 & -1 & 0 \\
1 & 0 & 1\end{array}\right)   \label{C2a}
\end{equation}
However, the latter one gives rise to an unstable
eigenvalue pair, $\lambda =\pm 2\epsilon $ (to leading order),
therefore it is not considered further. The former tile has a pair
of $\lambda =\pm 2\sqrt{2}\epsilon i$, and a double one, $\lambda
=\pm 2\epsilon i$, hence stable X-configurations (to be called 2a)
can be constructed at finite $\epsilon $ as arrays of such tiles
with alternating signs.

Other 
%relevant 
two-leg configurations are Y-shaped and ``skew-X"
ones, composed, respectively, of the 
%following 
tiles:
\begin{equation}
\left(
\begin{array}{ccc}
1 & 0 & -1 \\
0 & 0 & 0 \\
0 & 0 & 1\end{array}\right) ~\mathrm{and~}\left(
\begin{array}{ccc}
0 & -1 & 1 \\
0 & 1 & 0 \\
1 & -1 & 0\end{array}\right) .  \label{C2bc}
\end{equation}
Both these tiles are stable, with respective eigenvalues
$\lambda =\pm \sqrt{2}\epsilon i$ and $\lambda =\pm \sqrt{6}\epsilon i$
for the former, and $\lambda =\pm 0.874\sqrt{\epsilon }i$,
$\lambda =\pm 1.663\sqrt{\epsilon }i$,
$\lambda =\pm 2.882\sqrt{\epsilon }i$ and $\lambda =\pm
2.690\sqrt{\epsilon }i$ for the latter. Stable sign-alternate ENSs
assembled of them are termed 2b and 2c, respectively.

Three-leg ENSs can be built of other (generally, denser filled) \emph{stable}
tiles, such as
\begin{equation}
\left(
\begin{array}{ccc}
1 & -1 & 1 \\
-1 & 1 & -1 \\
1 & -1 & 1\end{array}\right) ~\mathrm{and~}\left(
\begin{array}{ccc}
1 & -1 & 1 \\
0 & 1 & 0 \\
-1 & 0 & 0\end{array}\right) .  \label{C3}
\end{equation}
The former tile possesses three pairs of double eigenvalues,
$\pm \sqrt{2\epsilon }i$, $\pm \sqrt{6\epsilon }i$, and $\pm \sqrt{8\epsilon }i$, and
two pairs of single ones, $\pm 2\sqrt{\epsilon }i$ and $\pm \sqrt{12\epsilon
}i$. The latter has a double eigenvalue $\pm \sqrt{2\epsilon }i$ and a
single one $\pm \sqrt{8\epsilon }i$. Three-leg configurations constructed of
these tiles with alternating signs are called 3a and 3b, see below. In fact,
the latter one will be a Z-shaped array.

Finally, we consider two configurations as a proof-of-principle of
the extension of the ENS\ concept to the 3D space. In particular,
augmenting the first tile of (\ref{C3}) by two $-1$ sites, adjacent
to middle $1$ along the third direction, we fabricate a ``stone".
The stone is stable, with single eigenvalues 
$\lambda =\pm 1.248\sqrt{\epsilon }i$, 
$\lambda =\pm 2\sqrt{\epsilon }i$,
$\lambda =\pm \sqrt{6\epsilon }i$, 
$\lambda =\pm 2.763\sqrt{\epsilon }i$ and 
$\lambda =\pm 3.848\sqrt{\epsilon }i$, and double and triple ones, 
$\lambda =\pm \sqrt{8\epsilon }i$ and 
$\lambda =\pm \sqrt{2\epsilon }i$, respectively. These stones
will be used to build a stable 3D pattern (again, with sign
alternations), called 4a in our nomenclature.

%Perhaps 
The simplest stable stone 
has a ``diamond"-like shape, 
%is shaped as a ``diamond", 
i.e., it consists of six AC-filled sites surrounding, as nearest
neighbors, an empty one. 
%While 
Although some configurations of this type
are unstable \cite{us_yuri}, we have found a stable one, 
%that we have found to be stable
carrying a phase distribution that corresponds to a quadrupole in the
plane, with phases $\pi /2$ and $3\pi /2 $ lent to the two
out-of-plane sites. This stone has triple and single eigenvalues,
$\lambda =\pm 2\epsilon i$ and $\lambda =\pm 4\epsilon i$,
respectively, and a higher-order one, $\mathcal{O}(\epsilon ^{2})$. 
Augmenting stones of this type by alternating phase pulses, we
will build a stable 3D structure labeled 4b.

\textit{Numerical results}. In Figs. \ref{Fig1}-\ref{Fig4} we
present, in a unified format, results of the numerical
continuation of various ENS configurations from the AC limit.
Figure \ref{Fig1} shows configuration 1 (a straight chain of
sign-alternating tiles). The norm of the configuration,
$N=\sum_{n}\phi _{n}^{2}$, and its maximum instability growth rate
are shown, as a function of $\epsilon $, in the top left and right
panel, respectively. In the 21x21 lattice used in our 2D
computation, the configuration becomes unstable for $\epsilon
>0.074$. The middle panels show typical examples of stable and
unstable configurations, and the bottom panels show the spectral
plane $(\lambda _{r},\lambda _{i})$ of their numerically found
(in)stability eigenvalues, $\lambda \equiv \lambda _{r}+i\lambda_{i}$, 
the instability being ushered by $\lambda _{r}\neq 0$.

Figure \ref{Fig2} presents three different varieties of the two-leg
configurations: 2a (X-waves, left panels), 2b (Y-waves, middle panels) and
2c (skewed X-waves, right panels). In addition, numerically computed
stability eigenvalues of the corresponding tiles, i.e., the first one in Eq.
(\ref{C2a}) and both in Eq. (\ref{C2bc}), are shown by solid blue lines for
comparison with the analytical predictions of the previous section (dashed
red lines), indicating good agreement between them. Configurations 2a, 2b,
and 2c become unstable at $\epsilon >0.202$, $\epsilon \geq
0.206$ and $\epsilon \geq 0.068$ respectively.

Similarly, Fig. \ref{Fig3} shows two three-leg configurations, 3a and 3b
(the latter one may be naturally called a Z-wave), and their stability
characteristics. These configurations become unstable at $\epsilon >0.053$
and $\epsilon \geq 0.075$, respectively.

Finally, figure \ref{Fig4} shows the continuation to $\epsilon
\neq 0$ of 3D configurations 4a and 4b (defined above), which are
stable, respectively, at $\epsilon \leq 0.043$ and $\epsilon \leq 0.075$.
Note that the corresponding numerically computed eigenvalues 
(for the respective stones) again
agree well with the analytical predictions.

\textit{Conclusion}. We have presented a systematic approach towards
constructing a variety of extended states on 2D and 3D nonlinear lattices.
They are composed of building blocks, namely, tiles (2D) or stones (3D) with
alternating signs, that are originally defined in the anti-continuum limit
(the sign alternation is necessary for stability of the patterns). Stability
intervals for the structures were revealed by numerical continuation of the
corresponding solution families to finite values of the coupling constant.
Examination of nonlinear evolution of unstable patterns is another relevant
problem, to be considered elsewhere.

Another interesting topic is a discrete model such as Eq.
(\ref{DNLS}), but with a negative intersite coupling in one
spatial direction (say, $x$); as mentioned above, this model can
be generated from Eq. (\ref{DNLS}) by means of the staggering
transformation applied in this direction, i.e., 
$u_{n_{x},n_{y},n_{z}}\equiv (-1)^{n_{x}}v_{n_{x},n_{y},n_{z}}$.
In the 2D version of such a model, one can immediately find two
exact \emph{linear} single-leg solutions oriented along the
diagonals, \textit{viz}., $u_{n_{x},n_{y}}=\exp (-|n_{x}\mp
n_{y}|)$. Their superposition can be used to construct exact
linear X-waves. A potential extension of the latter states into the 
nonlinear model would also be worth considering.

\end{document}